\documentclass[conference,a4paper]{APSIPA2020}
\usepackage{multirow}
\usepackage{graphicx}
\usepackage{amsmath}
\usepackage[psamsfonts]{amssymb}
\usepackage{amsxtra}
\usepackage{threeparttable}
\usepackage{url} 
\usepackage{hyperref}
\usepackage{bm}
\usepackage{siunitx}
\usepackage{subcaption}

\usepackage{tikz}
\usepackage{textcomp}
\usepackage{hyperref}
\usepackage{lipsum}
\newcommand\copyrighttext{%
	\footnotesize \textcopyright 
	Copyright 2020 IEEE. Published in the Asia-Pacific Signal and Information Processing Association Annual Summit and Conference 2020 (APSIPA 2020), scheduled for 7-10 December, 2020 in Auckland, New Zealand. Personal use of this material is permitted. However, permission to reprint/republish this material for advertising or promotional purposes or for creating new collective works for resale or redistribution to servers or lists, or to reuse any copyrighted component of this work in other works, must be obtained from the IEEE. Contact: Manager, Copyrights and Permissions / IEEE Service Center / 445 Hoes Lane / P.O. Box 1331 / Piscataway, NJ 08855-1331, USA. Telephone: + Intl. 908-562-3966.
	}
\newcommand\copyrightnotice{%
	\begin{tikzpicture}[remember picture,overlay]
	\node[anchor=south,yshift=10pt] at (current page.south) {\fbox{\parbox{\dimexpr\textwidth-\fboxsep-\fboxrule\relax}{\copyrighttext}}};
	\end{tikzpicture}%
}

\begin{document}

\title{A Study on More Realistic Room Simulation for Far-Field Keyword Spotting}

\author{%
\authorblockN{%
Eric Bezzam\authorrefmark{1}\authorrefmark{3},
Robin Scheibler\authorrefmark{2},
Cyril Cadoux\authorrefmark{3}, and
Thibault Gisselbrecht\authorrefmark{1}
}
\authorrefmark{1}
Sonos Inc., Paris, France \\
\authorrefmark{2}
Line Corporation, Tokyo, Japan \\
\authorrefmark{3}
\'Ecole Polytechnique F\'ed\'erale de Lausanne, Switzerland \\
\authorblockA{%
E-mail: eric.bezzam@epfl.ch}
}

\maketitle
\copyrightnotice
\thispagestyle{empty}

\begin{abstract}
  We investigate the impact of more realistic room simulation for training far-field keyword spotting systems without fine-tuning on in-domain data.
  To this end, we study the impact of incorporating the following factors in the room impulse response (RIR) generation: air absorption, surface- and frequency-dependent coefficients of real materials, and stochastic ray tracing.
  Through an ablation study, a wake word task is used to measure the impact of these factors in comparison with a ground-truth set of measured RIRs.
  On a hold-out set of re-recordings under clean and noisy far-field conditions, we demonstrate up to $35.8\%$ relative improvement over the commonly-used (single absorption coefficient) image source method.
  Source code is made available in the \emph{Pyroomacoustics} package, allowing others to incorporate these techniques in their work.
\end{abstract}

\section{Introduction}
\label{sec:intro}

Deep-learning approaches are the state-of-the-art when it comes to keyword spotting (KWS) and automatic speech recognition (ASR)~\cite{hinton2012deep}. 
Their recent success can be attributed to the availability of large datasets, improved computing resources, and a cocktail of deep-learning heuristics that have been tried-and-tested by an active research community. Each data point and its reproducibility contribute to a better understanding of the best practices for speech recognition training.

Given an appropriate dataset, deep neural networks (DNN) are able to learn internal representations that are relatively stable with respect to variables independent of the desired output. 
For example in KWS and ASR tasks, the speaker's pronunciation should not impact the understanding of what the speaker is saying.
Robustness to speaker pronunciation, e.g.\ gender and accents, 
is typically handled by collecting a sufficiently varied dataset, such as Librispeech for ASR~\cite{Panayotov2015}, Google's speech commands for KWS~\cite{Warden2018} and the ``Hey Snips'' dataset for wake word (WW) detection~\cite{Coucke2019}. Additional variations, e.g.\ speed and pitch, can be simulated to further augment the original dataset and expose the model to more variety during training~\cite{Ko2013, Jaitly2013}. 

With the recent proliferation of smart speaker devices, there has been a growing interest and need for robust, far-field recognition, namely speech in the presence of reverberation and various types of noise. 
As collecting and labelling such in-domain data for training can be difficult and time-consuming, multi-condition training (MCT), i.e.\ simulating various acoustic environments, offers an attractive alternative~\cite{Gao2020, Ko2017, Kim2017}.
Far-field settings are simulated by convolving clean anechoic recordings with room impulse responses (RIR), which can be either simulated or measured. 


In~\cite{Ko2017, Kim2017}, the authors augment their dataset through MCT with RIRs that are generated with the image source method (ISM)~\cite{Allen1979}. Both~\cite{Ko2017} and~\cite{Kim2017} show gains on a hold-out set of real recordings.
The authors of~\cite{Tang2019} investigate a room simulation technique that uses stochastic ray tracing (SRT), in order to incorporate factors not taken into account by ISM, i.e.\ diffuse reflections and late reverberation.
Their results demonstrate an improved performance on KWS and ASR tasks when using SRT instead of ISM.

The use of measured RIRs for MCT is preferable as it is difficult for simulation to fully capture the complexity of real rooms, e.g.\ the exact shape, materials, and furniture inside, and how sound propagates within them, e.g.\ low frequency wave effects. 
Although not as time-consuming as collecting in-domain data, gathering a sufficiently varied dataset of measured RIRs (in terms of number of rooms) is equally difficult to obtain. Therefore, most of the past work has relied on simulated RIRs or a mix of simulated and measured RIRs~\cite{Ko2017}.
Moreover, there have been numerous studies on the differences between using simulated and measured RIRs, with several attempts to close the performance gap: point-source noises~\cite{Ko2017}, directional sources~\cite{Ravanelli2016}, and artificially mimicking low-frequency wave effects~\cite{Tang2020}.



SpecAugment~\cite{Park2019} is another data augmentation technique that has shown promising results for ASR. Within the context of far-field recognition, its effect is unclear, with worse results shown in~\cite{Park2020} when used in conjunction with MCT. As our focus is on far-field recognition and the use of MCT to generalize to these conditions, we do not use SpecAugment in our study.

While only SRT with a single absorption coefficient is used in~\cite{Tang2019}, we propose to employ a \emph{hybrid} approach by complementing ISM with SRT~\cite{Schroder2011}. We also incorporate air absorption and surface- and frequency-dependent absorption coefficients of real materials in an attempt to bring simulation closer to reality.
The goal of this paper is two-fold:
\begin{itemize}
	\item Study the impact of incorporating the above factors into ISM with an ablation study on a WW detection task. 
	To the best of our knowledge, there is no such ablation study within the context of speech recognition applications.
	\item Demonstrate and make available such techniques through the \emph{Pyroomacoustics} package~\cite{Scheibler2018}.\footnote{
		\href{https://github.com/LCAV/pyroomacoustics}{\texttt{github.com/LCAV/pyroomacoustics}}
	}
	\footnote{
		Room impulse response generation code for this paper:\\
		\href{https://github.com/ebezzam/room-simulation}{\texttt{github.com/ebezzam/room-simulation}}
	}
\end{itemize}

In our ablation study, we use a dataset of measured RIRs, namely the BUT Speech@FIT Reverb Database (ReverbDB)~\cite{Szoke2019}, to quantify the gap between currently-used room simulation techniques for KWS and ASR, i.e.\ ISM and SRT, and the ideal scenario of having a dataset consisting of only measured RIRs. Although ReverbDB does not contain the amount of rooms and RIRs to train a model robust to the large variety that would be encountered by a typical smart speaker device, it is of sufficient size to serve as an oracle dataset for our controlled setup and contains the metadata of each room to simulate the various proposed factors.

The paper is organized as such: Section~\ref{sec:related} presents ISM and SRT; Section~\ref{sec:contribution} describes how the proposed factors, i.e.\ the hybrid approach, air absorption, and surface- and frequency- dependent absorption coefficients of real materials, are incorporated; Section~\ref{sec:setup} and~\ref{sec:results} detail our experimental setup and results; and Section~\ref{sec:conclusion} contains concluding remarks.

\section{Modeling room impulse responses}
\label{sec:related}

For $I$ sources, we can model far-field speech as:
\begin{equation}
\label{eq:sim}
x[n] = \sum_{i=1}^I g_i \left(p_i[n] \ast h_{i}[n]\right),
\end{equation}
where $\{p_i[n]\}_{i=1}^I$ are the $I$ source signals (target and noise), each convolved with an RIR $\{h_{i}[n]\}_{i=1}^I$ from the $i$-th source to the microphone and scaled with a constant $\{g_i\}_{i=1}^I$.

\subsection{Image source method}

The ISM technique is a popular approach for generating RIRs, mainly due to its simplicity~\cite{Ko2017, Kim2017, Allen1979}. 
It models reflections off walls as virtual sources outside of the room, but at a distance corresponding to the length of the reflection path within the room. 
With ISM, we can write the RIR for the $i$-th source as:
\begin{equation}
\label{eq:ism}
h_{i}[n] = \sum_{\bm{s} \in V(\bm{m}, \bm{s}_i)} \frac{ R(\bm{m}, \bm{s}) }{4 \pi \|\bm{m} - \bm{s}\|} \delta_{LP} \left(n - F_s\frac{\|\bm{m} - \bm{s}\|}{c}\right),
\end{equation}
where $\bm{m}$ is the position of the microphone, $\bm{s}_i$ is the position of the $i$-th source, $V(\bm{m}, \bm{s}_i)$ is the set of visible image sources for the reflections between $\bm{m}$ and $\bm{s}_i$, $\delta_{LP}$ is a windowed sinc function~\cite{Scheibler2018}, $F_s$ is the sample rate,   $c$ is the speed of sound, and $R(\bm{m}, \bm{s})$ is the accumulated reflection coefficient between $\bm{m}$ and $\bm{s}$:
\begin{equation}
\label{eq:abs}
R(\bm{m}, \bm{s}) = \prod_{w \in W(\bm{m}, \bm{s})} \sqrt{1 - \alpha_w^2},
\end{equation}
where $W(\bm{m}, \bm{s})$ is the list of surfaces in the reflection path between $\bm{m}$ and $\bm{s}$, and $\alpha_w$ is the absorption coefficient of surface $w$.

A maximum reflection order $M$, i.e.\ how many wall reflections to simulate, is selected to build $ V(\bm{m}, \bm{s}_i)$. For arbitrary-shaped rooms, the simulation complexity is exponential with regards to the reflection order: $\mathcal{O}(N^M)$ where $N$ is the number of reflective surfaces. For cuboid-shaped (i.e.\ shoebox) rooms, the complexity simplifies to $\mathcal{O}(M^3)$  due to the symmetry of the room~\cite{Allen1979}. 

\subsection{Stochastic ray tracing}

\begin{figure}[t]
	\centering
	\includegraphics[width=0.6\linewidth]{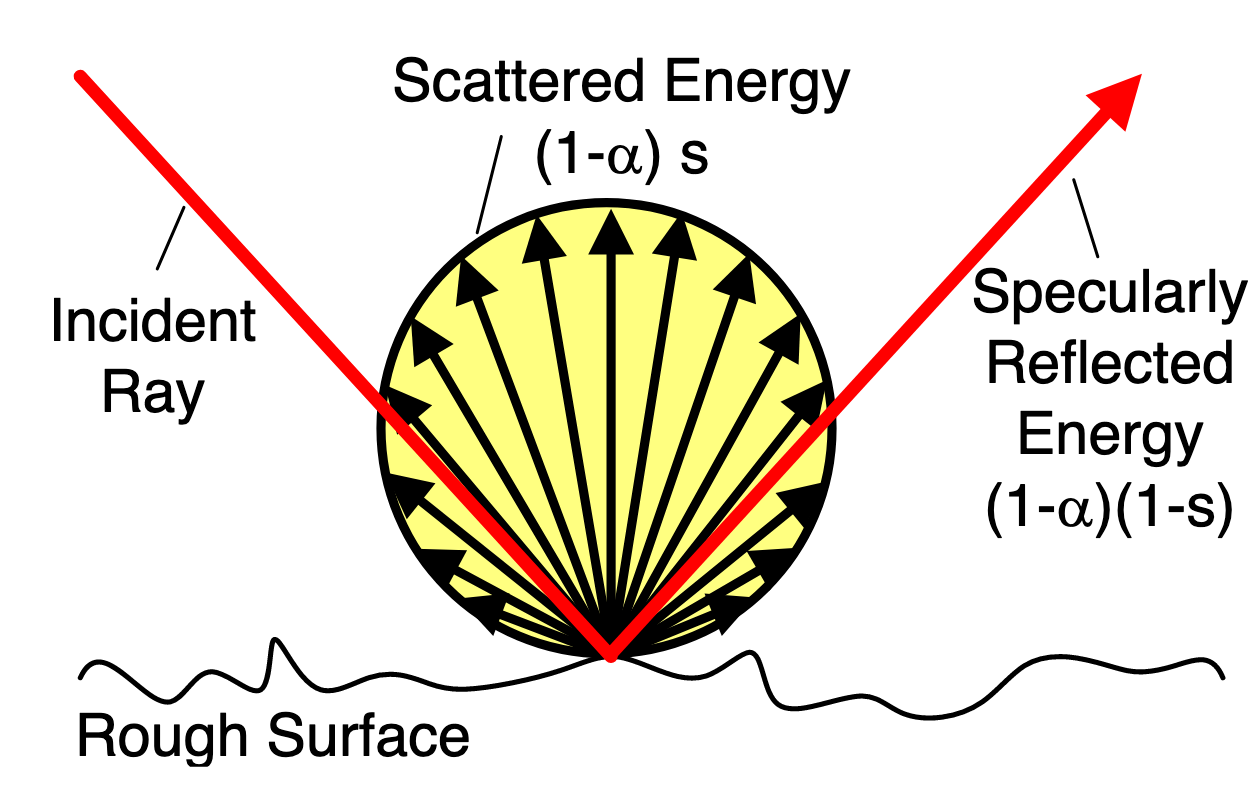}
	\caption{Illustration of the separation of reflected energy into scattered and specular components~\cite{Schroder2011}. $\alpha$ and $s$ are the surface's absorption and scattering coefficients.}
	\label{fig:reflection_energy}
\end{figure}

Up to a desired order $M$, the ISM technique is able to perfectly model \emph{specular} reflections, i.e.\ reflections that have the same outgoing angle as the incident angle to a surface. This modeling is accurate only as long as the wavelength of the sound is small relative to the size of the reflector. When this is not the case, we have \emph{diffuse} reflections that scatter in all directions~\cite{Schroder2011}. Fig.~\ref{fig:reflection_energy} illustrates this separation into specular and scattered energy.

Furthermore, the reflection order $M$ limits the ability of ISM to capture the late reverberation of a room. 
Increasing the ISM order can capture more reflections in order to better model the late reverberation, but diffuse reflections are still not taken into account. With SRT, we can model these diffuse reflections and the late reverberation~\cite{Tang2019, Schroder2011}.
From each point source, a large number of rays are emitted and traced until the energy of each ray falls below a certain threshold.
Each microphone is modeled as a receiver volume, with specular rays that intersect with it contributing to the resulting RIR. The received rays are logged with an amplitude and timestamp in order to produce an energy histogram.

Unlike ISM, each reflection may emit scattered sound energy back to the receiver.
In this work, we employ the \emph{diffuse rain} technique~\cite{Schroder2011}, in which a secondary ray is emitted at each reflection point in the direction of the receiver. The energy of this secondary ray
can be written as:
\begin{equation}
E_{scat} = E_{in} \cdot (1-\alpha) \cdot s \cdot P_{Hit},
\end{equation}
where $E_{in}$ is the ray's incoming energy, $\alpha$ and $s$ are the surface's absorption and scattering coefficients respectively, and $P_{Hit}$ is the probability that the scattered energy reaches the receiver~\cite{Schroder2011}. The specular ray's energy is given by: 
\begin{equation}
E_{spec} = E_{in} \cdot (1-\alpha) \cdot (1 - s).
\end{equation}

In order to produce the RIR, the envelope of the resulting energy histogram is used to shape the envelope of a randomly generated sequence of Dirac deltas~\cite{Schroder2011}.

\section{Proposed simulation for training}
\label{sec:contribution}

In Section~\ref{sec:hybrid}, we describe the hybrid room simulation technique; and in Section~\ref{sec:additional_factor}, we discuss how to incorporate air absorption and surface- and frequency-dependent coefficients.

\subsection{Hybrid approach}
\label{sec:hybrid}

Due to its stochastic nature, we are not guaranteed to receive all specular reflections with SRT. A hybrid approach can capture these reflections up to a desired order $M$:
\begin{enumerate}
	\item Apply ISM of order $M$ for specular reflections: $ h_{i}^{ISM}[n] $.
	\item Apply SRT for diffuse reflections and late reverberation: $ h_{i}^{SRT}[n] $.
	\item Add the two simulations for the hybrid RIR:
	\begin{equation}
	h_{i}[n] =  h_{i}^{ISM}[n] + h_{i}^{SRT}[n].
	\end{equation}
\end{enumerate}
When performing SRT, specular reflections which are at or below $M$ should be neglected, as to avoid counting the same reflections twice. Moreover, the energy levels between ISM and SRT must be balanced at the start of simulation~\cite{Schroder2011}.

\subsection{Additional factors}
\label{sec:additional_factor}

Air absorption is incorporated by introducing a $e^{-\gamma r}$ factor to the RIRs, where $r$ is the total distance travelled and $\gamma$ is an air attenuation coefficient. This coefficient depends on temperature, humidity, and frequency~\cite{Kuttruff2000}.

Current approaches to room simulation for speech recognition employ a single absorption coefficient for the entire room~\cite{Ko2017, Kim2017}. In reality, absorption properties depend on the material of the surface and on the frequency. This dependence can be taken into account by performing ISM and SRT with unique absorption coefficients for each surface with~\ref{eq:abs} and on separate octave bands. An appropriately designed filter bank $\{\phi_f[n]\}_{f=1}^F$ is used to combine the frequency-dependent RIRs at the end of simulation:
\begin{equation}
h_{i}[n] = \sum_{f=1}^F  \phi_f[n] \ast ( h_{i}^{ISM, f}[n] + h_{i}^{SRT, f}[n] ).
\end{equation}

\begin{figure}[t]
	\centering
	\includegraphics[width=1.0\linewidth]{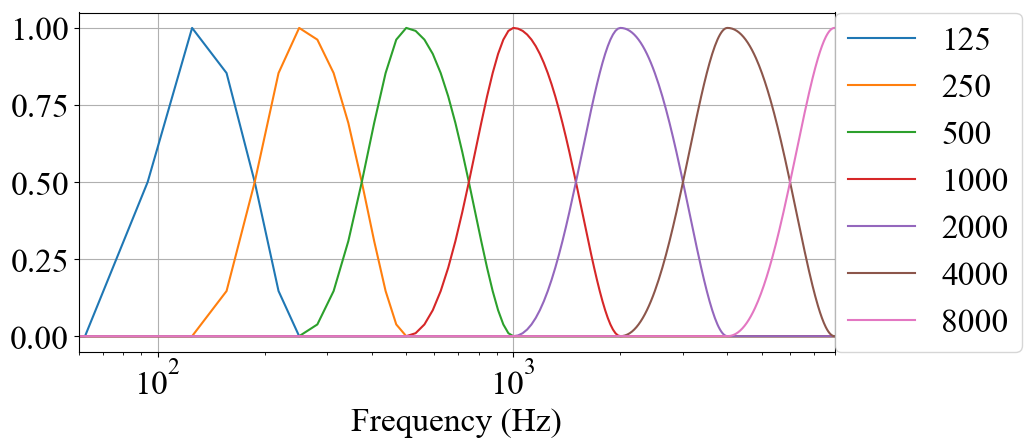}
	\caption{Third-octave filter bank to simulate frequency-dependent room impulse responses.}
	\label{fig:filter_banks}
\end{figure}

In this work, we use a third-octave filter bank, with $F=7$ FIR bandpass filters centered at $[125, 250, 500, 1000, 2000,$ $ 4000, 8000]$ \SI{}{\hertz}, as these are frequencies at which absorption and scattering coefficients of materials are typically measured~\cite{Vorlander2013}. The filters are designed using the window method with an FFT of $512$ and a raised cosine window. The frequency response of each filter can be seen in Fig.~\ref{fig:filter_banks}.

\section{Experimental setup}
\label{sec:setup}

A WW task is used for our ablation study as the smaller model size allows for quicker iterations and more trainings in order to perform a more in-depth analysis. Moreover, a binary task allows the use of DET (Detection Error Tradeoff) curves to analyze the performance over a wide range of operating points instead of a fixed point. 

%


Three types of datasets are used in our study:
\begin{enumerate}
	\item WW data (positive and negative utterances from the same speakers): ``Hey Snips'' dataset~\cite{Coucke2019}. 
	\item Measured RIRs: BUT Speech@FIT Reverb Database (ReverbDB)~\cite{Szoke2019}.
	\item Noise datasets: MUSAN~\cite{Snyder2015} for non-speech sounds and Librispeech~\cite{Panayotov2015} for interfering speech.
\end{enumerate}
We use ReverbDB as the oracle RIRs that we wish to emulate via simulation. The dataset contains essential metadata in order to study the impact of the factors we wish to investigate.
In Table~\ref{tab:data}, the train / dev / test set distribution (before augmentation) is detailed. 

\begin{table}[b]
	\caption{Original dataset distribution.}
	\begin{center}
		
		\begin{tabular}{ c c c c }
			\hline
			& train & dev & test \\ 
			\hline
			\# ``Hey Snips''  & $5876$ & $2588$  & $2504$\\ 
			\# negative & $45344$  & $20821$   & $20321$\\ 
			\# rooms & $7$  & $2$  &  $1$ \\ 
			\# noise samples & $15504$  & $2299$  &  $2087$ \\ 
			\hline
		\end{tabular}
		
	\end{center}
	\label{tab:data}
\end{table}

In Section~\ref{sec:sim}, we describe how ReverbDB is used to simulate RIRs with the different simulation factors. In Section~\ref{sec:train_dev} and Section~\ref{sec:test}, we describe the augmented train, dev, and test sets for our study.

\subsection{RIR simulation}
\label{sec:sim}

For each room simulation variation, we emulate the RIRs from the ReverbDB dataset using the provided metadata. The split in Table~\ref{tab:data} corresponds to $1032$ RIRs for the train set and $280$ RIRs for the dev set. For all simulations, the same room dimensions, microphone positions, and speaker positions are used, as specified in the ReverbDB metadata.
\begin{enumerate}
	\item \textbf{Baseline}: a single absorption coefficient is used for all frequencies and all walls of a shoebox room as in~\cite{Ko2017, Kim2017, Tang2019}. Using Eyring's equation as in~\cite{Kim2017}, this coefficient is computed from the estimated RT60\footnote{Time for the sound to decay by $60$dB.} provided in the ReverbDB metadata for each microphone-speaker pair.
	\item \textbf{AIR}: the measured temperature for each room in ReverbDB is used to set the appropriate frequency-dependent air absorption coefficients as specified in~\cite{Vorlander2013}. This temperature is also used to set the speed of sound.\footnote{$\text{speed of sound} = 331.4 + 0.6 \times \text{temperature}$}
	\item \textbf{MAT}: using the name of the floor, ceiling, and wall materials provided for each room in ReverbDB, we identify the corresponding frequency-dependent coefficients in~\cite{Vorlander2013} for each wall. These values are used instead of the single absorption coefficient. It is important to note that these absorption coefficients are not provided in the ReverbDB metadata but are taken from a separate resource. Therefore, the values may not represent the actual absorption coefficients of the room walls. Moreover, we do not take into account furniture in the room.
	\item \textbf{MAT AIR}: Finally, we use both AIR and MAT.
\end{enumerate}
All four of the above variations are applied for ISM, SRT, and the hybrid approach (HYB) for our ablation study. An ISM order of 17 is used for both ISM and HYB. For the ray tracing approaches, we use the percentage of furniture covering in the ReverbDB metadata as a single scattering coefficient, even for the frequency-dependent simulation in MAT. All of the simulation variations are generated using \emph{Pyroomacoustics}. As a direct comparison to~\cite{Tang2019}, we also perform Baseline with their simulator;\footnote{
	\href{https://github.com/RoyJames/pygsound}{\texttt{github.com/RoyJames/pygsound}}
} incorporating the other factors is not possible with their current library.

\subsection{Train and dev sets}
\label{sec:train_dev}

The original train and dev sets are augmented as such, with the WW data (positive and negative utterances) as the target speaker:
\begin{enumerate}
	\item A room is sampled at random, from which we sample a microphone, one speaker for the target speaker, and ($75\%$ of the time) 1 or 2 speaker(s) to serve as point-source noise(s) as in~\cite{Ko2017}.
	\item Pitch and speed variations are applied to the target speaker as in~\cite{Ko2013}.
	\item Using~\ref{eq:sim}, the target and sampled noise sources are simulated in the corresponding room with an SNR sampled as in~\cite{Kim2017}. 
\end{enumerate}
Each sample in the original train set is augmented $16$ times with different samplings in order to produce the dataset that is used for training the WW detector. This $16$x augmentation is performed for all $13$ room simulation variations described in Section~\ref{sec:sim} and the ReverbDB dataset, resulting in a total of $14$ types of room simulation.

\subsection{Test set}
\label{sec:test}

For evaluating the different room simulation approaches, we use re-recordings in order to observe how the simulation approaches generalize to real far-field scenarios. The original test set in Table~\ref{tab:data} is re-recorded in five conditions within a typical office setting: clean, $\SI{5}{\decibel}$ non-speech, $\SI{5}{\decibel}$ speech, $\SI{2}{\decibel}$ non-speech, and $\SI{2}{\decibel}$ speech. The target speaker is placed $\SI{3}{\meter}$ away. For the noisy conditions, a single speaker is placed $\SI{2}{\meter}$ away at a $45\si{\degree}$ angle with respect to the target.

\subsection{Model architecture and training}

The WW model employed in this study follows the architecture and training hyper-parameters from~\cite{Coucke2019}. The model takes inspiration from WaveNet~\cite{Oord2016}, as it uses dilated convolutions to capture long temporal patterns, gated activation units to control the propagation of information to the next layer, and skip connections to speed up convergence and avoid vanishing gradient issues. In total, there are $24$ layers, corresponding to a receptive field of \SI{1.83}{\second}. The input acoustic features are $20$-dimensional log-Mel filterbank energies (LFBEs), extracted every \SI{10}{\milli\second} over a window of \SI{25}{\milli\second}. Xavier initialization is used, and the Adam optimization method is used with a learning rate of $10^{-3}$.

As the focus of this paper is on room simulation techniques, we refer the interested reader to~\cite{Coucke2019} for a more detailed description on model architecture and training hyper-parameters, and an ablation study on the importance of gating and skip connections.

For each simulation approach, $10$ models are trained with different seeds in order to prevent random effects due to initialization or sampling from affecting the analysis of our results (by averaging the performance over the $10$ models).

\begin{table*}[t]
	\caption{(Top half) average relative change in false rejection rate with respect to using measured RIRs (ReverbDB), for three false alarms per day. Higher is better, with values close to $0$ indicating that the simulation technique is on par with using measured RIRs. (Bottom half) average relative change with respect to using ISM. Confidence intervals are for three standard deviations.}
	\begin{center}
		\begin{tabular}{l >{\centering\arraybackslash}p{2.5cm}   >{\centering\arraybackslash}p{2.5cm}  >{\centering\arraybackslash}p{2.5cm} >{\centering\arraybackslash}p{2.5cm}}
			\hline
			& \multicolumn{1}{c}{\textbf{Baseline}} & \multicolumn{1}{c}{\textbf{AIR}} & \multicolumn{1}{c}{\textbf{MAT}} & \multicolumn{1}{c}{\textbf{MAT AIR}} \\ 
			\hline
			\textbf{ISM} &     $-55.1 \pm 1.61\%$   &    $-16.7 \pm 0.453\%$    &    $0.78 \pm 0.291 \%$   &   $-14.3 \pm 0.709 \%$      \\ 
			\textbf{SRT~\cite{Tang2019}} &   $-32.2 \pm 1.02\%$  &  $-$  & $-$ &  $-$       \\ 
			\textbf{SRT} &   $-18.0 \pm 0.403\%$  &  $-14.6 \pm 0.219\%$  & $-52.2 \pm 1.33\%$ &  $-25.1 \pm 0.670\%$         \\ 
			\textbf{HYB} &  $-10.4 \pm 0.284\%$  &  $-5.02 \pm 0.321\%$  &  $-40.3 \pm 0.850\%$ & $-34.9 \pm 0.859\%$     \\ 
			\hline 
			\hline
			\textbf{ISM} &     $-$   &    $24.3\pm 0.830\%$    &    $35.8\pm 1.60\%$   &   $25.3 \pm 1.14\%$      \\ 
			\textbf{SRT~\cite{Tang2019}} &   $15.2\pm 0.920\%$  &  $-$  & $-$ &  $-$       \\ 
			\textbf{SRT} &   $24.1\pm 1.38\%$  &  $27.1\pm 1.82\%$  & $3.92\pm 0.817\%$ &  $20.0 \pm 1.60\%$         \\ 
			\textbf{HYB} &  $28.8\pm 1.80\%$  &  $33.0\pm 2.11\%$  &  $10.5\pm 0.775\%$ & $13.9 \pm 0.977\%$     \\ \hline
		\end{tabular}
	\end{center}
	\label{tab:frr}
\end{table*}

\begin{figure*}[t!]%
	\centering
	\begin{subfigure}{.68\columnwidth}
		\includegraphics[width=\columnwidth]{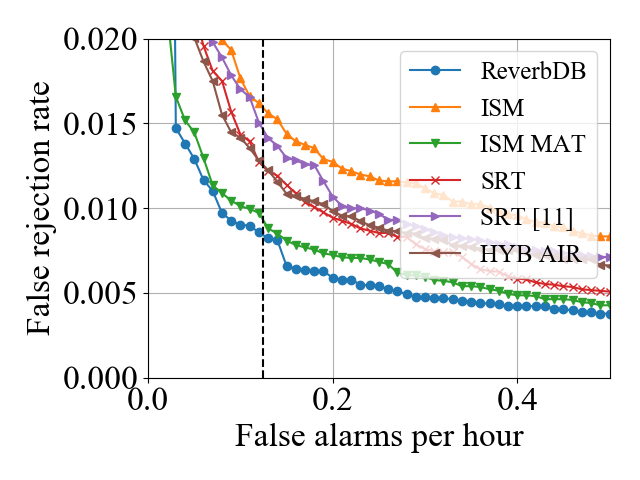}%
		\caption{clean}%
		\label{clean}%
	\end{subfigure}\hfill%
	\begin{subfigure}{.68\columnwidth}
		\includegraphics[width=\columnwidth]{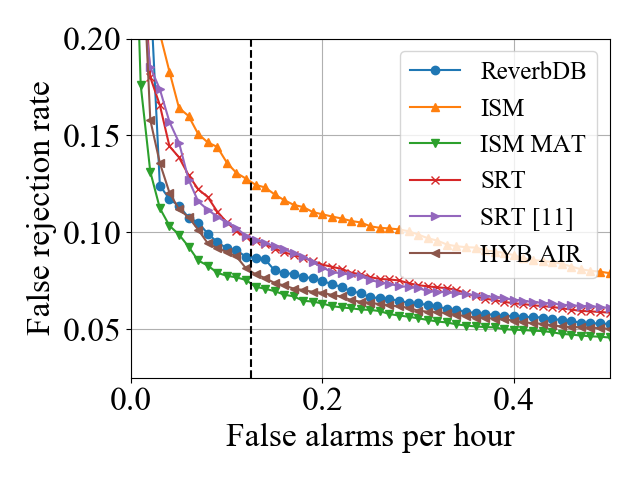}%
		\caption{non-speech $\SI{5}{\decibel}$}%
		\label{non-speech5}%
	\end{subfigure}\hfill%
	\begin{subfigure}{.68\columnwidth}
		\includegraphics[width=\columnwidth]{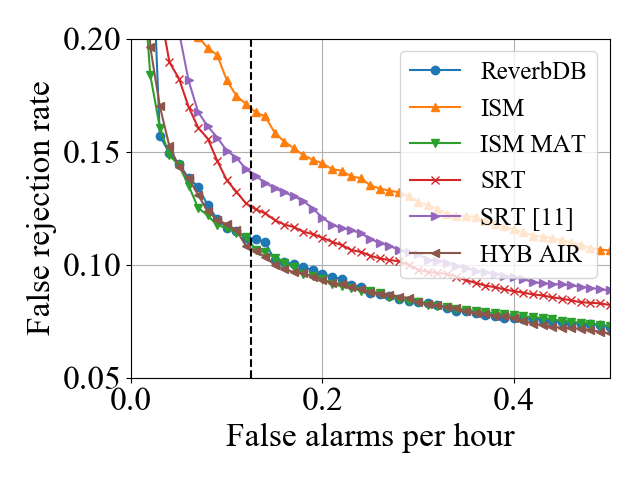}%
		\caption{speech $\SI{5}{\decibel}$}%
		\label{speech5}%
	\end{subfigure}\hfill
	\begin{subfigure}{.68\columnwidth}
		\includegraphics[width=\columnwidth]{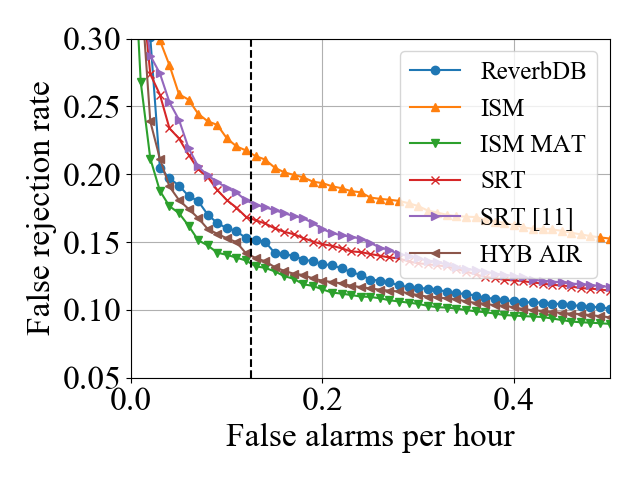}%
		\caption{non-speech $\SI{2}{\decibel}$}%
		\label{non-speech2}%
	\end{subfigure}
	\begin{subfigure}{.68\columnwidth}
		\includegraphics[width=\columnwidth]{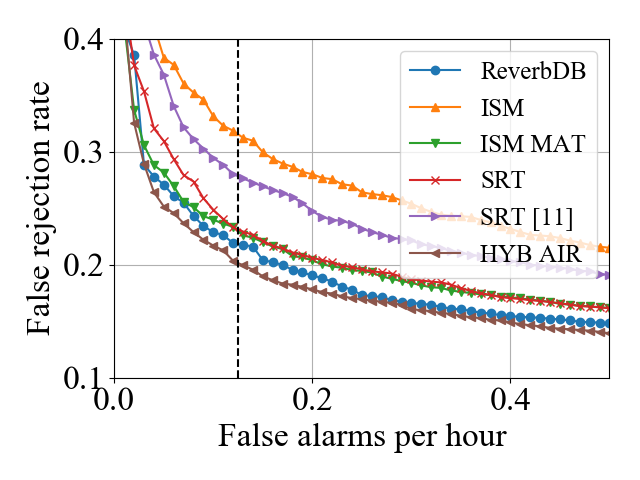}%
		\caption{speech $\SI{2}{\decibel}$}%
		\label{speech2}%
	\end{subfigure}%
	
	\caption{Detection error tradeoff curves comparing room simulation techniques (averaged over $10$ random seeds) for a WW (``Hey Snips'') detection task: false alarm per hour ($x$-axis) vs.~false rejection rate ($y$-axis). Vertical, black dotted lines indicate the performance for three false alarms per day. 
		For all conditions, the target is placed $\SI{3}{\meter}$ from the device and for noisy conditions, the interfering sound is placed $\SI{2}{\meter}$ at a $45\si{\degree}$ angle with respect to the target. Note that the $y$-axis is scaled differently for each condition to better visualize the difference between simulation approaches.
	}
	\label{fig:det}
\end{figure*}

\begin{figure*}[t]%
	\centering
	\begin{subfigure}{1.0\columnwidth}
		\includegraphics[width=\columnwidth]{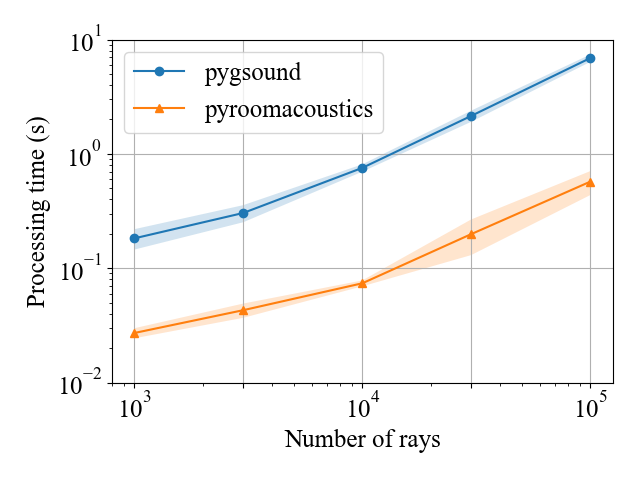}%
		\caption{Varying number of emitted rays (for diffuse reflections).}%
		\label{nrays}%
	\end{subfigure}
	\begin{subfigure}{1.0\columnwidth}
		\includegraphics[width=\columnwidth]{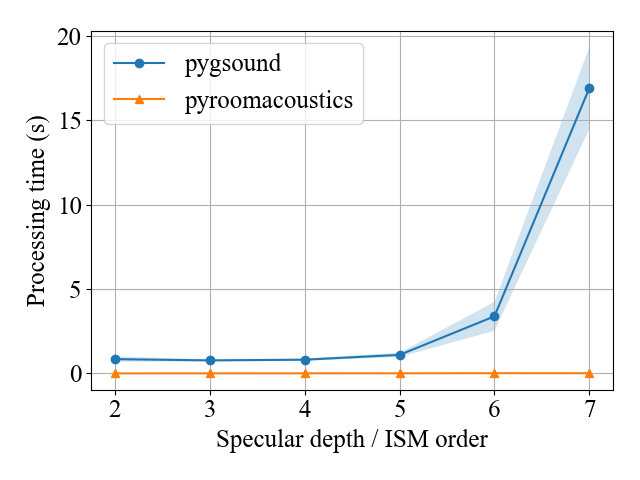}%
		\caption{Varying specular depth / ISM order.}%
		\label{ism}%
	\end{subfigure}
	\caption{Average runtime (over $100$ trials) for computing an RIR within a $8\si{\meter} \times 9\si{\meter} \times  3\si{\meter}$ room with an RT60 of $0.5\si{\second}$ and a scattering coefficient of $0.5$. Computations are performed on a MacBook Pro 2014 with a 2.6 GHz Dual-Core. In~\ref{nrays}, the number of emitted rays is varied with a fixed specular depth / ISM order of $3$; 
		in~\ref{ism} the specular depth is increased for \emph{pygsound} while the ISM order is increased for \emph{Pyroomacoustics} for a fixed number of emitted rays ($1e4$). Confidence intervals are for one standard deviation.
	}
	\label{fig:profile}
\end{figure*}

\section{Results and analysis}
\label{sec:results}

The top half of Table~\ref{tab:frr} shows the average relative change in false rejection rate for each simulation approach with respect to using measured RIRs (ReverbDB). This value is computed by averaging over the relative changes for the five re-recorded far-field conditions. Fig.\@~\ref{fig:det} shows the DET curves for a select few simulation techniques.

A significant gap can be observed between the commonly-used ISM approach and using the measured RIRs of ReverbDB: $55.1\%$ drop in average relative performance. This gap can be seen for each condition in Fig.\@~\ref{fig:det}.

From the Baseline column in Table~\ref{tab:frr}, we observe that the proposed HYB approach outperforms currently-used techniques: ISM~\cite{Ko2017, Kim2017} and SRT~\cite{Tang2019}. 
As presented in~\cite{Tang2019}, we observe an improvement of SRT over ISM. The further increase in performance from SRT to HYB seems to indicate the need of early specular reflections, which are not guaranteed with SRT alone. The difference between our implementation of SRT and~\cite{Tang2019} may arise from different simulation parameters and internal implementations.\footnote{See~\cite{Schissler2011} for implementation details of the simulator in~\cite{Tang2019}.}

From Table~\ref{tab:frr}, we see that AIR benefits all baseline techniques; perhaps the induced energy decay results in more realistic RIRs. On the other hand, the impact of MAT is not clear. For ISM, we observe a significant improvement, effectively closing the gap with ReverbDB. For the ray tracing techniques, we see a large degradation in performance with respect to the baseline approaches. A potential reason for this is the scattering coefficient used in simulation. As there is no metadata on the scattering properties of the rooms in the ReverbDB, we use the furniture coverage as a scattering coefficient for all surfaces and frequencies, which could be very far from reality. This modeling choice would not impact ISM simulation as scattering is not taken into account. Furthermore, the MAT AIR column seems to indicate we do not get additive gains from incorporating both proposed factors.

The effect of MAT highlights a fundamental issue that arises when trying to match simulation to real world RIRs, namely the appropriate choice of simulation parameters and their distribution. It could be that the name of the wall materials, as provided in the ReverbDB metadata, is not enough information to perform a surface- and frequency- dependent emulation of a room.
As the absorption coefficients used for MAT are deduced from a separate dataset~\cite{Vorlander2013}, they could be very different from the coefficients of the wall (and furniture) materials in the corresponding room of ReverbDB.

Despite the unclear results on incorporating MAT, there is a clear indication that adding ray tracing and AIR is beneficial. Moreover, all proposed simulation factors display an improvement over ISM (bottom half of Table~\ref{tab:frr}).

\subsection{Profiling RIR simulators}

In Fig.\@~\ref{fig:profile}, we compare the computational time between our implementation of SRT in \emph{Pyroomacoustics} with that of \emph{pygsound}, namely the simulator used in~\cite{Tang2019}. For an increasing number of emitted ray in Fig.\@~\ref{nrays}, we observe that \emph{Pyroomacoustics} is roughly an order of magnitude faster than \emph{pygsound}. For an increasing specular depth in Fig.\@~\ref{ism}, we see an exponential growth in \emph{pygsound}. 
Specular depth is roughly equivalent to the ISM order, namely how many specular wall reflections are included in the RIR modeling. 
This seems to suggest that \emph{pygsound} does not take into account the symmetry of shoebox-shaped rooms.

\section{Conclusions and future work}
\label{sec:conclusion}

In this paper, we investigated the impact of more realistic room simulations on far-field wake word (WW) training without fine tuning on in-domain, recorded data. To this end, we quantified the gap between an oracle set of measured RIRs (ReverbDB) and the commonly-used ISM approach for room simulation. Through an ablation study, we studied the effect of incorporating additional simulation factors, i.e.\@ stochastic ray tracing for diffuse and late reverberation, air attenuation (AIR), and surface- and frequency-dependent absorption coefficients (MAT).

On a hold-out set of re-recordings under clean and noisy far-field conditions, we observe a $28.8\%$ average relative improvement to ISM when complementing it with ray tracing (HYB). Moreover, we find that incorporating AIR benefits all simulation techniques (a further improvement to $33.0\%$ when used with HYB). The impact of MAT is unclear. With ISM, we found the gap between simulated and measured RIRs to be effectively closed. With ray tracing techniques, we found a degradation in performance, which may be due to improper modeling of scattering properties, for which we did not have metadata about in ReverbDB.

As future work, we would like to investigate the impact of such factors on other far-field speech recognition tasks, e.g.\@ command detection and ASR. Finally, the source code for generating the RIRs is made available in the \emph{Pyroomacoustics} package. We hope this will aid other researchers and engineers in incorporating such techniques in their work.

The purpose of this work was to quantify the gap between various simulation techniques and the ideal case of using measured RIRs. However, in order to have a fair comparison with the measured RIRs, we were limited in terms of the number of rooms and RIRs we could simulate. However, to use the techniques presented in this paper and achieve the variety of rooms needed for far-field KWS and ASR tasks, one could envision a room generation scheme similar to one in~\cite{Kim2017}; where instead of sampling an RT60, variations in air absorption attenuation and different frequency-dependent materials for each wall could be sampled.\footnote{In the repository for this paper, we have included all the frequency-dependent coefficients from~\cite{Vorlander2013}:\href{https://github.com/ebezzam/room-simulation/blob/master/materials.py}{\texttt{room-simulation/materials.py}} }

\bibliographystyle{IEEEtran}
\bibliography{refs_mendeley_apsipa}

\end{document}